\documentclass[epj,twocolumn]{webofc}
\usepackage[varg]{txfonts}   
\usepackage{graphicx}
\usepackage{epstopdf, epsfig}
\usepackage{xcolor}
\usepackage{bm}
\usepackage{physics}
\usepackage{tabularx}
\usepackage{wasysym}
\woctitle{Powders \& Grains 2021}
\begin{document}
\title{Designing non-segregating granular mixtures}
%
%

\author{\firstname{Yifei} \lastname{Duan}\inst{1} \and
        \firstname{Paul B.} \lastname{Umbanhowar}\inst{2} \and
        \firstname{Richard M.} \lastname{Lueptow}\inst{1,2}\thanks{\email{r-lueptow@northwestern.edu}}
}

\institute{Department of Chemical and Biological Engineering, Northwestern University, Evanston, IL 60208, USA
\and
Department of Mechanical Engineering, Northwestern University, Evanston, IL 60208, USA
          }

\abstract{%
In bidisperse particle mixtures varying in size or density alone, large particles rise (driven by percolation) and heavy particles sink (driven by buoyancy).
When the two particle species differ from each other in both size and density, the two segregation mechanisms either enhance (large/light and small/heavy) or oppose (large/heavy and small/light) each other.
In the latter case, an equilibrium condition exists in which the two mechanisms balance and the particles no longer segregate. 
This leads to a methodology to design non-segregating particle mixtures by specifying particle size ratio, density ratio, and mixture concentration to achieve the equilibrium condition. 
Using DEM simulations of quasi-2D bounded heap flow, we show that segregation is significantly reduced for particle mixtures near the equilibrium condition. In addition, the rise-sink transition for a range of particle size and density ratios matches the predictions of the combined size and density segregation model.
}
\maketitle
\section{Introduction}
\label{introduction}
Flowing granular materials can segregate due to differences in constituent particle size or density.
In dense granular flows of size-disperse particles having the same density (S-system), large particles tend to rise as small particles fall through voids, a segregation mechanism known as percolation \cite{savage1988particle}.
For density-disperse mixtures of equal size particles (D-system), segregation is driven by a buoyant force mechanism in which heavy particles sink and light particles rise \cite{khakhar1997radial}.
When particles differ from each other in both size and density (SD-system), the two segregation mechanisms interact, resulting in more complicated segregation behavior. 
Though size and density differences can reinforce each other, e.g., in mixtures of large light  particles and small heavy particles, we are interested here in the situation where the two segregation mechanisms oppose each other, e.g., in mixtures of large heavy  particles and small light particles, as it has the potential to reduce segregation compared to the corresponding S- or D-system.

\begin{figure}
\centering
\includegraphics[width=0.5\textwidth,clip]{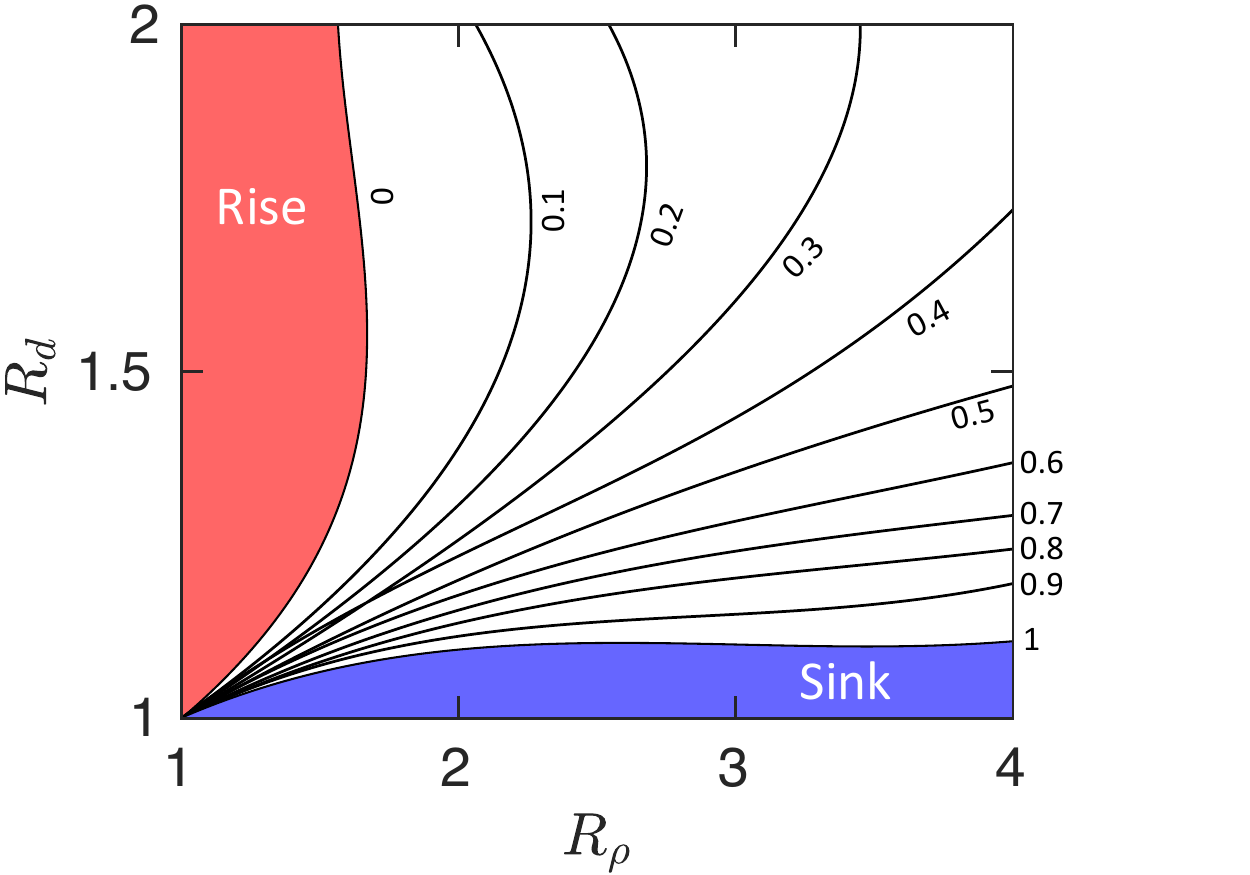}
\caption{
Equilibrium (no segregation) concentration of large particles, $c_{l,eq}$, for a bidisperse mixture vs. particle size and density ratios. 
Along iso-concentration curves of $c_{l,eq}$, particle mixtures remain mixed for the corresponding $R_d$ and $R_\rho$. 
For $R_d$ and $R_\rho$ in the colored regions, no equilibrium concentration exists and segregation is uni-directional.
}
\label{fig-1}       
\end{figure}

Previous studies have shown that the tendency of spherical particles to sink or rise in a bidisperse mixture can be characterized by the ratios of large to small particle diameter, $R_d=d_l/d_s$ (subscript $l$ for large
particles and $s$ for small particles regardless of their densities), and density, $R_\rho=\rho_l/\rho_s$, along with the mixture volume concentration $c_l$ (or equivalently $c_s$, as $c_l$+$c_s$=1) \cite{gray2005theory,tunuguntla2017comparing,fan2014modelling,jones2018asymmetric}.
Unlike size or density segregation alone, where the particles alway rise or sink regardless of the mixture concentration, the segregation direction in an SD-system can be concentration dependent \cite{alonso1991optimum,felix2004evidence,gray2015particle}.
To capture the concentration dependent segregation direction in SD-systems, a semi-empirical segregation velocity model that depends linearly on the local shear rate and quadratically on the species concentration has been proposed \cite{duan2020modelling}.
This model has two empirical coefficients that are functions of $R_d$ and $R_\rho$. 
Concentration profiles predicted by incorporating this segregation velocity model into a continuum advection-diffusion-segregation transport model match DEM simulation results well for a wide range of $R_d$ and $R_\rho$.
One result of the model is the prediction shown in Fig.~1 of the large particle "equilibrium concentration," $c_{l,eq}$, at which the size-related percolation and the density-related buoyancy offset one another such that the segregation flux of the two species is zero.  In this paper, we use this result to demonstrate how non-segregating granular mixtures can be designed based on appropriate choices of $R_d$, $R_\rho$, and $c_l$.

\section{Non-segregating mixtures}

A practical concern for many industrial situations is assuring that a mixture of two particle species remains mixed despite their differences in size and density.
Fig.~1, which is based on the combined size and density segregation model \cite{duan2020modelling}, not only provides the equilibrium concentration at which mixtures of particles with specified combinations of size and density ratios remain mixed, but also indicates the segregation direction of each particle species. If the mixture concentration of large particles for specific values of  $R_d$ and $R_\rho$ is greater than the corresponding equilibrium concentration curve $c_{l,eq}$ large particles rise, whereas if the concentration is less than the $c_{l,eq}$ curve large particles sink.
Segregation direction is independent of concentration, or uni-directional (shaded regions), for ($R_d$, $R_\rho$) combinations to the left of the $c_{l,eq}=0$ curve (large particles rise) and to the right of the $c_{l,eq}=1$ curve (large particles sink).

In many practical situations the material of each particle species is fixed, thereby fixing the density ratio, but the 
species sizes can be altered as desired.
This opens the possibility of specifying particle sizes to avoid segregation at desired mixture concentrations.
Here we demonstrate the potential for designing non-segregating granular mixtures in this way.

\section{Simulations}


Using our in-house DEM code \cite{isner2020axis} running on CUDA-enabled GPUs, we numerically simulate combined size and density segregation of bidisperse mixtures in a single-sided quasi-2D bounded heap to validate the non-segregating equilibrium conditions shown in Fig.\,1.  
Simulation setup and segregation examples are shown in Fig.\,2.
The heap is confined by two parallel plates in the spanwise direction with a gap thickness of 1.5\,cm.
The horizontal heap width is 40\,cm.
The bottom wall is inclined at $28^\circ$ from horizontal, roughly matching the repose angle in order to reduce computation time (fewer particles required).
The standard linear spring-dashpot model \cite{cundall1979discrete} is used to resolve particle-particle and particle-wall contacts using a friction coefficient of 0.5, a restitution coefficient of 0.2, and a binary collision time of 0.5\,ms, for all simulations.

A stream of well-mixed particles with a feed concentration of large particles, $c_l$,  is fed into the system at a constant 2-dimensional rate of $q=20\,$cm$^2$/s.
For the heap flows considered here, segregation occurs in a thin flowing layer near the free surface, which is about 5$d_l$ in depth.
Fig.~2 shows resulting segregation for $ c_l=0.1$, $R_d=1.5$, and three different values of $R_\rho$.  
For $R_\rho=1$ large particles segregate upward and deposit downstream near the endwall, indicating that percolation dominates. 
Particles remain relatively mixed for $R_\rho=2$, as the two segregation mechanisms are nearly balanced. 
Segregation reverses for $R_\rho=3$ as buoyancy dominates percolation.

\begin{figure}
\centering
\includegraphics[width=0.5\textwidth,clip]{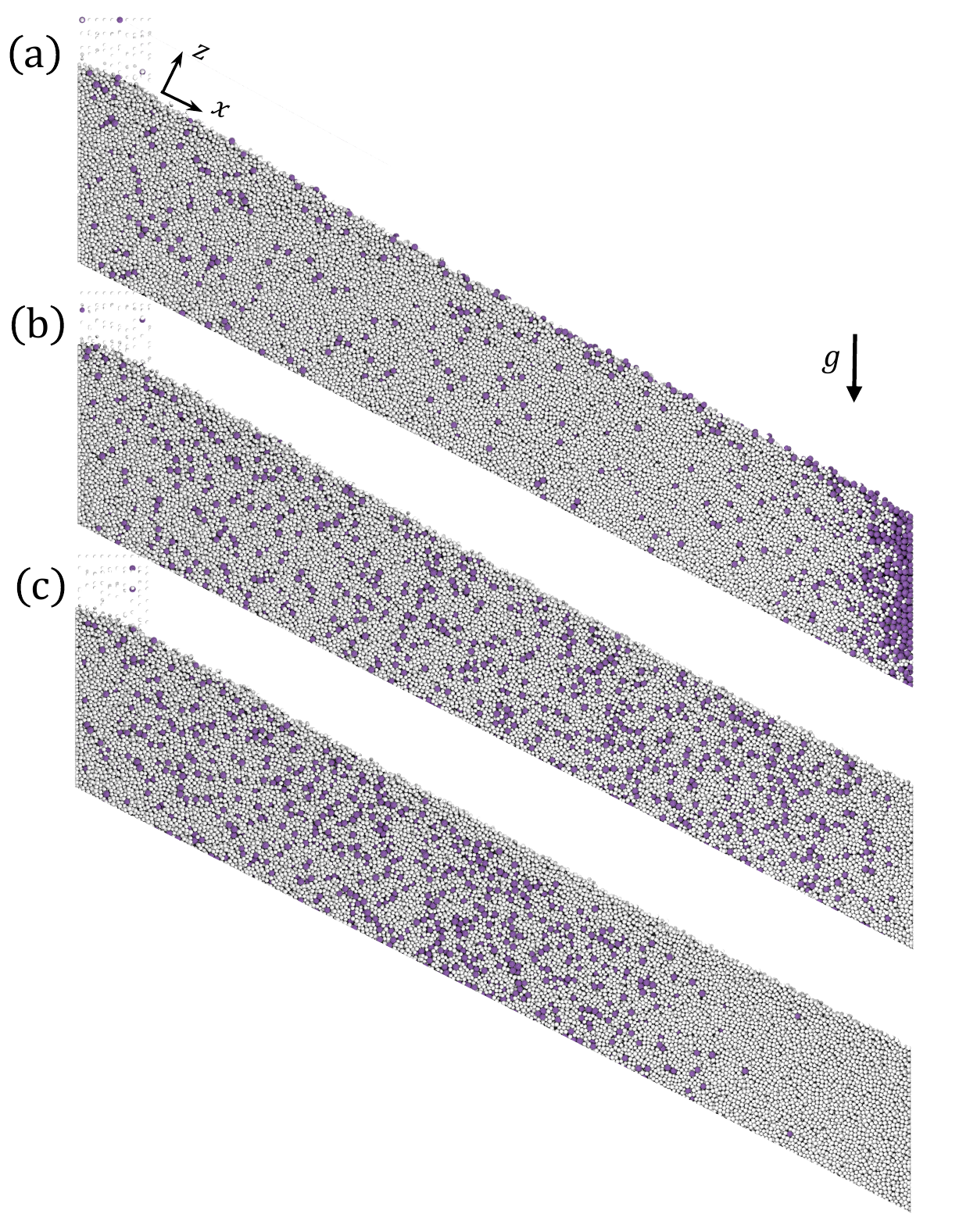}
\caption{
Heap flow segregation for $ c_l=0.1$ and size ratio $R_d=1.5$ showing reversal in large particle segregation direction for density ratio $R_\rho$ of (a) 1, (b) 2, and (c) 3. 
Large particles are \textcolor{black}{purple} ($d_l=3\,$mm, $\rho_l=R_\rho\rho_s$) and small particles are \textcolor{black}{gray} ($d_s=2\,$mm, $\rho_s=1$g/cm$^3$).
}
\label{fig-0}       
\end{figure}
 
 \begin{figure}
\centering
\includegraphics[width=0.5\textwidth,clip]{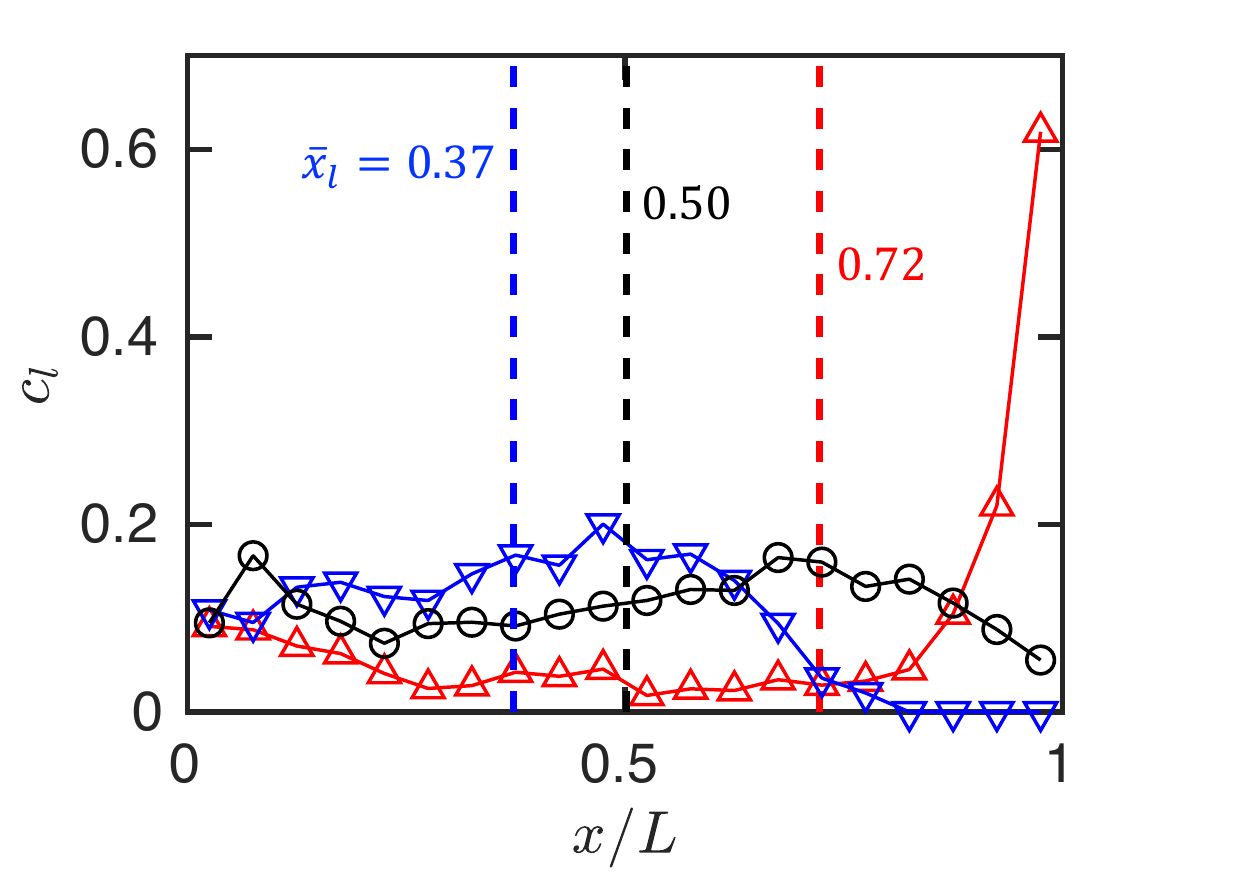}
\caption{
Streamwise concentration profiles for large particles deposited on the heap.
Symbols represent local concentration for the three cases in Fig.~2 with $R_\rho$ of 1 ($\textcolor{red}{\triangle}$), 2 ($\Circle$), and 3 (\textcolor{blue}{$\triangledown$}).
Dashed lines indicate the mean streamwise position (centroid) of large particles $\bar x_l$.
}
\label{com}       
\end{figure}

To quantify the degree of segregation, we calculate the dimensionless large-particle-concentration-based mean streamwise position (centroid),
\begin{equation}
\bar x_l=\frac{1}{L}\frac{\sum^{P}_{p=1}x_p c_{l,p}}{\sum^{P}_{p=1} c_{l,p}},
\label{xl}
\end{equation}
where $P=20$ is the number of uniform width bins for calculating $c_{l}$ at different streamwise positions, and $c_{l}$ is the local depth-averaged volume concentration for particles deposited on the heap below the flowing layer.
Fig.~3 plots $c_{l}$ profiles for the three cases in Fig.~2 and indicates the corresponding values for $\bar x_l$.  
For $R_\rho=1$ (red curve), the concentration of large particles is highest near $x/L\approx1$ as they deposit on the downstream portion of the heap along with a few small particles, consistent with the observation in Fig.~2(a).
For this case $\bar x_l=0.72$.
The situation reverses for $R_\rho=3$ (blue curve), where a high concentration of large particles deposit on the upstream portion of the heap, and pure small particles deposit on the downstream portion, as observed in Fig.~2(c), resulting in $\bar x_l=0.37$.
For an intermediate case  where size and density effects are nearly balanced, large particles deposit nearly uniformly on the heap and $c_l$ remains relatively constant (black curve), resulting in $\bar x_l=0.50$.

An ideal non-segregating case with constant $c_{l}$ has $\bar x_l=0.50$.
This is consistent with the measured value of $\bar x_l$ for the weak segregation case (black curve) in Fig.~3. For the two segregating cases, $\bar x_l$ deviates from 0.5 as expected since for perfect segregation, $\bar x_l=1-c_l/2$ for rising large particles and $c_l/2$ for sinking large particles. 
Consequently, a simple scalar for the offset of the mean streamwise position, $\bar x_l-0.5$, is used to characterize both the segregation direction and the degree of segregation. 

\begin{figure}
\centering
\includegraphics[width=0.5\textwidth,clip]{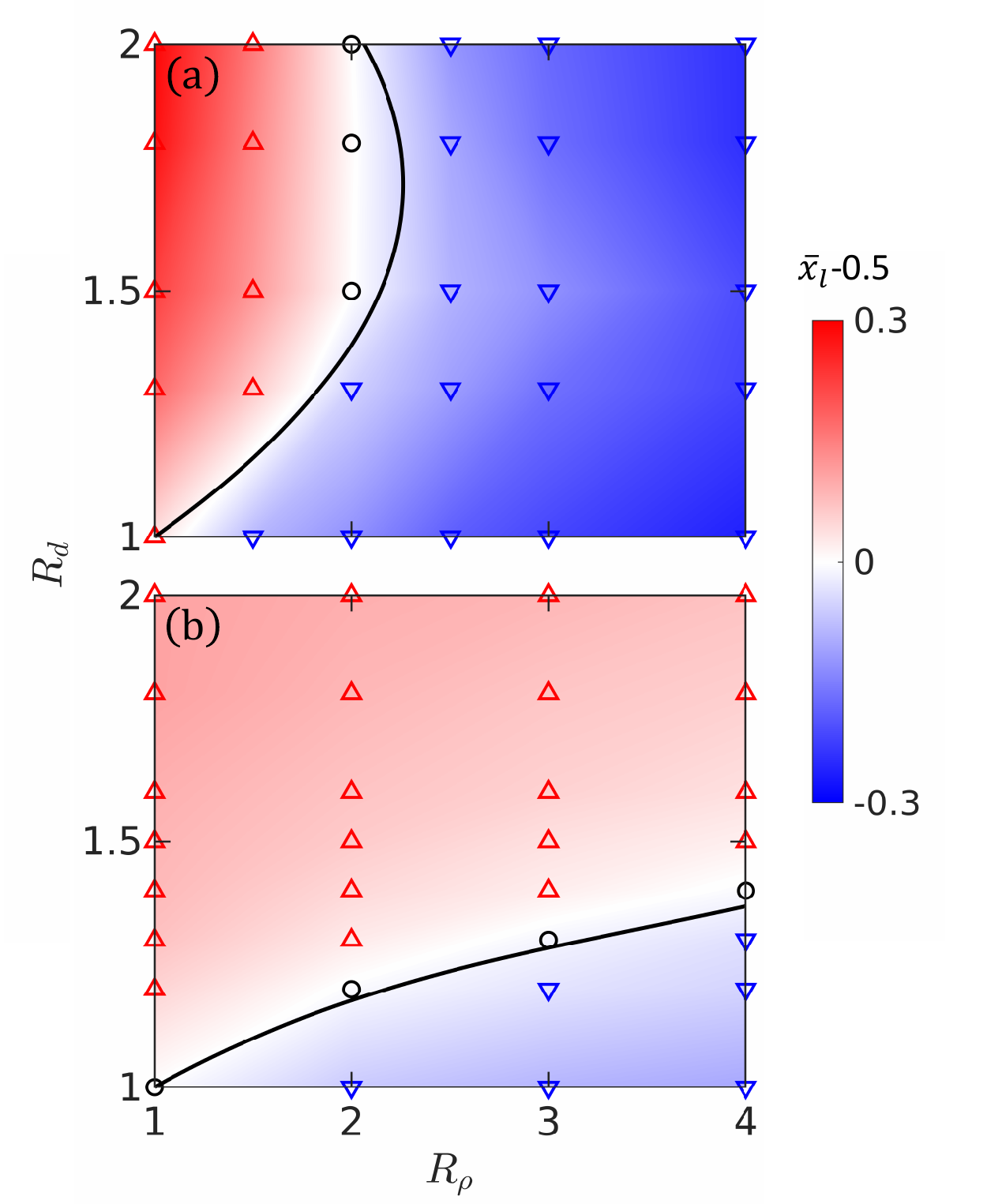}
\caption{
Offset of  the mean streamwise position for large particles, $\bar x_l-0.5$, (color contours) for feed concentration $c_l$ of (a) 0.1 and (b) 0.6.  Symbols indicate offsets greater than 0 ($\textcolor{red}{\triangle}$, rising large particles), approximately zero, i.e., 0$\pm$0.015 ($\Circle$, no segregation), and less than 0 ($\textcolor{blue}{\triangledown}$, sinking large particles).  Color contours are interpolated from data points, each corresponding to a different DEM simulation.
}
\label{c01}       
\end{figure}

Fig.~4 demonstrates the rise-sink transition in heap flow simulations (data points) for multiple $R_d$ and $R_\rho$ combinations, i.e., 30 for $c_l=0.1$ in Fig.~4(a) and 32 for $c_l=0.6$ in Fig.~4(b). 
Red and blue shading corresponds to the degree that $\bar x_l$ deviates from the mixed state value of $\bar x_l=0.50$ with white corresponding to particles remaining mixed.
The predicted equilibrium curve for the specified value of $c_l$ from Fig.\,1 is represented by the black curve here and corresponds closely to the white region, indicating that the particles remain mixed at that feed concentration.  In other words, the $c_{l,eq}$ curve in the $R_d$,$R_\rho$ plane along which particles are predicted to remain mixed at that concentration from Fig. 1 (black curve) corresponds closely to the combinations of $R_d$ and $R_\rho$ along which particles remain mixed for that feed concentration (white region where $\bar x_l$=0.5).
The small deviation of the black curve from the white region could be the result of many factors including the resolution of data points in the ($R_d$,$R_\rho$) space and fluctuations in the local concentration similar to those observed for the weakly segregating case (black curve) shown in Fig.~3.

Particle mixtures with ($R_d,R_\rho$) not on the equilibrium curve in Fig.~4 segregate as expected.  If the ($R_d,R_\rho$) pair falls on an equilibrium curve in Fig.~1 with $c_{l,eq}<c_l$, then the feed concentration is too high to maintain equilibrium, and the large particles will rise resulting in $\bar x_l>0.5$.  If the ($R_d,R_\rho$) pair corresponds to a higher equilibrium concentration curve in Fig. 1, then the feed concentration is too low to maintain equilibrium, and the large particles will sink resulting in $\bar x_l<0.5$.  An equivalent alternative explanation is that for $c_l = c_{l,eq}$, moving below the equilibrium curve in the $R_d,R_\rho$-plane corresponds to smaller size ratios and larger density ratios, indicating that large particles will sink; moving above the equilibrium curve corresponds to larger size ratios and smaller density ratios, indicating that large particles will rise. The actual deposition
of particles indicated by the colors in Fig. 4 for specific values of $c_l$ match this prediction.

\section{Summary}

For particle mixtures varying simultaneously in size and density, the two corresponding segregation mechanisms (percolation and buoyancy, respectively) interact with each other resulting in segregation behavior significantly different from size or density segregation alone.
In particular, mixtures of large heavy and small light particles can have an equilibrium concentration at which the two segregation mechanisms are balanced and the net segregation flux is zero.
This leads to a methodology in which a particle system can be designed to prevent segregation by specifying the optimal combination of particle size ratio $R_d$, density ratio $R_\rho$, and mixture concentration $c_l$. 

The overlap of the equilibrium curves in Fig.\,4 with the white non-segregating region determined from DEM simulations  for both feed concentrations that were considered not only demonstrates the accuracy of the equilibrium conditions predicted in Fig.\,1 but also further validates the combined size and density segregation model described in Ref.~\cite{duan2020modelling}.

In many situations $R_d$ and $R_\rho$ are fixed but local concentration $c_l$ can vary over time and space, which may make the equilibrium condition difficult to maintain.
However, for heap flows like those in Fig.\,2 where the segregation time scale is short, particles quickly deposit onto the heap.
In such cases, particle mixtures designed to be at the equilibrium concentration, as shown in Fig. 1, exhibit minimal segregation, as indicated by the simulation results in Fig. 4. 
Further work is needed to investigate such non-segregating mixtures in other flow geometries. 
For example, for particles in rotating tumblers a small fluctuation in  local concentration could, with sufficient time, result in a global instability, i.e., segregation. 
Nevertheless, these results demonstrate the potential of specifying particular combinations of particle properties and concentration to promote mixing and minimize segregation.

\section*{Acknowledgements}
We thank Yi Fan and John Hecht for valuable discussions.
This material is based upon work supported by the National Science Foundation under Grant No. CBET-1929265.

%

\begin{thebibliography}{}
%
%


\bibitem{savage1988particle}
S.B. Savage and C.K.K. Lun, J. Fluid Mech., \textbf{189}, 311 (1988).
\bibitem{khakhar1997radial}
D.V. Khakhar, J.J. McCarthy, and J.M. Ottino, Phys. Fluids, \textbf{9}, 3600 (1997).
\bibitem{gray2005theory}
J.M.N.T. Gray and A.R. Thornton, Proc. R. Soc. A, \textbf{461}, 1447 (2005).
\bibitem{tunuguntla2017comparing}
D.R. Tunuguntla, T. Weinhart, and A.R. Thornton, Compu. Part. Mech., \textbf{4}, 387 (2017).
\bibitem{fan2014modelling}
Y. Fan, C.P. Schlick, P.B. Umbanhowar, J.M. Ottino, and R.M. Lueptow, J. Fluid Mech., \textbf{741}, 252 (2014).
\bibitem{jones2018asymmetric}
R.P. Jones, J.M. Ottino, P.B. Umbanhowar, and R.M. Lueptow, Phys. Rev. Fluids, \textbf{3}, 094304 (2018).
\bibitem{alonso1991optimum}
M. Alonso, M. Satoh, and K. Miyanami, Powder Technol., \textbf{68}, 145 (1991).
\bibitem{felix2004evidence}
G. F{\'e}lix and N. Thomas, Phys. Rev. E \textbf{70}, 051307 (2004).
\bibitem{gray2015particle}
J.M.N.T. Gray and C. Ancey, J. Fluid Mech., \textbf{779}, 622 (2015).
\bibitem{duan2020modelling}
Y. Duan, P. B. Umbanhowar, J.M. Ottino, and R.M. Lueptow, arXiv:{2011.09018}, (2020).
\bibitem{isner2020axis}
A.B. Isner, P.B. Umbanhowar, J.M. Ottino, and J.M. Lueptow, Chem. Eng. Sci., \textbf{217}, 115505 (2020).
\bibitem{cundall1979discrete}
P.A. Cundall and O.D.L. Strack, G{\'e}otechnique, \textbf{29}, 47 (1979).


\end{thebibliography}
%
%

\end{document}